\documentclass[3p,times,procedia]{elsarticle}
\usepackage{nupha_ecrc}

\volume{00}

\firstpage{1}

\journalname{Nuclear Physics A}

\runauth{}

\jid{nupha}

\jnltitlelogo{Nuclear Physics A}

\usepackage{amssymb}
\usepackage{caption}
\usepackage{subcaption}
\usepackage{lineno}

\usepackage[figuresright]{rotating}

\begin{document}
\begin{frontmatter}



\dochead{}

\title{Addressing the hypertriton lifetime puzzle with ALICE at the LHC}


\author{Stefano Trogolo \\ \normalsize{on behalf of the ALICE Collaboration}}

\address{Universit\`a di Torino, Dip. Fisica and INFN, Sez. Torino - Via Giuria 1, 10125 Torino, Italy}

\begin{abstract}
The ALICE Collaboration collected a large data sample of Pb-Pb collisions at $\sqrt{s_{\rm{NN}}}$ = 5.02 TeV in 2015 at the Large Hadron Collider (LHC) and the excellent particle identification (PID) capabilities allow for the detection of rarely produced (anti-)hypernuclei.
In particular, the (anti-)hypertriton, $^3_{\Lambda}$H, which is a bound state of a proton, a neutron and a $\Lambda$, is the lightest known hypernucleus. 

The results on the $^3_{\Lambda}$H production are compared with the predictions from a model based on coalescence approach and from statistical-thermal models to investigate the production mechanisms in heavy-ion collisions. 
Emphasis will also be put on the latest and more precise determination of the $^3_{\Lambda}$H lifetime. 
\end{abstract}


\end{frontmatter}


\section{Introduction}
\label{intro}
In ultra-relativistic heavy-ion collisions at the LHC, a state of matter called Quark-Gluon Plasma (QGP) is created. Among the particles produced in these collisions, (anti-)hypernuclei are of special interest since the production mechanism of these loosely bound states is not clear. Hypernuclei are nuclei where a nucleon is replaced with a hyperon. The hypertriton ($^3_{\Lambda}$H), a bound state of a proton, a neutron and a $\Lambda$, is the lightest known hypernucleus. Its mass is 2.99116 $\pm$0.00005 GeV/$c^2$ \cite{Davis} and the decay channels accessible to the ALICE apparatus are the mesonic ones with charged daughters, namely \mbox{$^3_{\Lambda}$H $\rightarrow ^3$He + $\pi^-$}, \mbox{$^3_{\Lambda}$H $\rightarrow $ d + p + $\pi^-$} and their charge conjugates. 

Two different groups of models are used to describe the production of (anti-)(hyper-)nuclei in such collisions: the thermal model \cite{PLB697:2011} and the coalescence of hadrons \cite{Coal_1, Coal_2}.
The thermal model allows us to calculate the production yields of hadrons (d\textit{N}/d\textit{y}) created in the fireball, under the assumption of thermal equilibrium, when the chemical freeze-out temperature (T$_{chem}$) is reached and the inelastic collisions cease.
The coalescence approach assumes that (anti-)baryons, if close enough in the phase-space at kinetic freeze-out can form a multi-baryon state, when elastic collisions cease.

Moreover, the $^3_{\Lambda}$H lifetime determination is one of the open points in hypernuclear physics, since recent measurements showed values off from the expected value between 232 and 256 ps \cite{Congletone:1992jp, Kamada:1997rv}. A new and more precise lifetime measurement of the ALICE Collaboration is presented in this report.

The results were obtained by analyzing the data sample of Pb--Pb collisions at $\sqrt{s_{\mathrm{NN}}}$ = 5.02 TeV and exploiting the mesonic weak decay into two charged particles.
The decay daughters were identified via specific energy-loss in the Time Projection Chamber (TPC) and topological cuts were applied to reconstruct and select the secondary vertices. The signal extraction was performed by means of a fit to the invariant mass distribution with a function that includes signal (Gaussian) and background (2$^{nd}$ degree polynomial) contributions.

\section{Production}
\label{prod}
\begin{figure}[!h]
\centering
\begin{subfigure}[h]{.5\textwidth}
  \centering
  \includegraphics[scale=0.1]{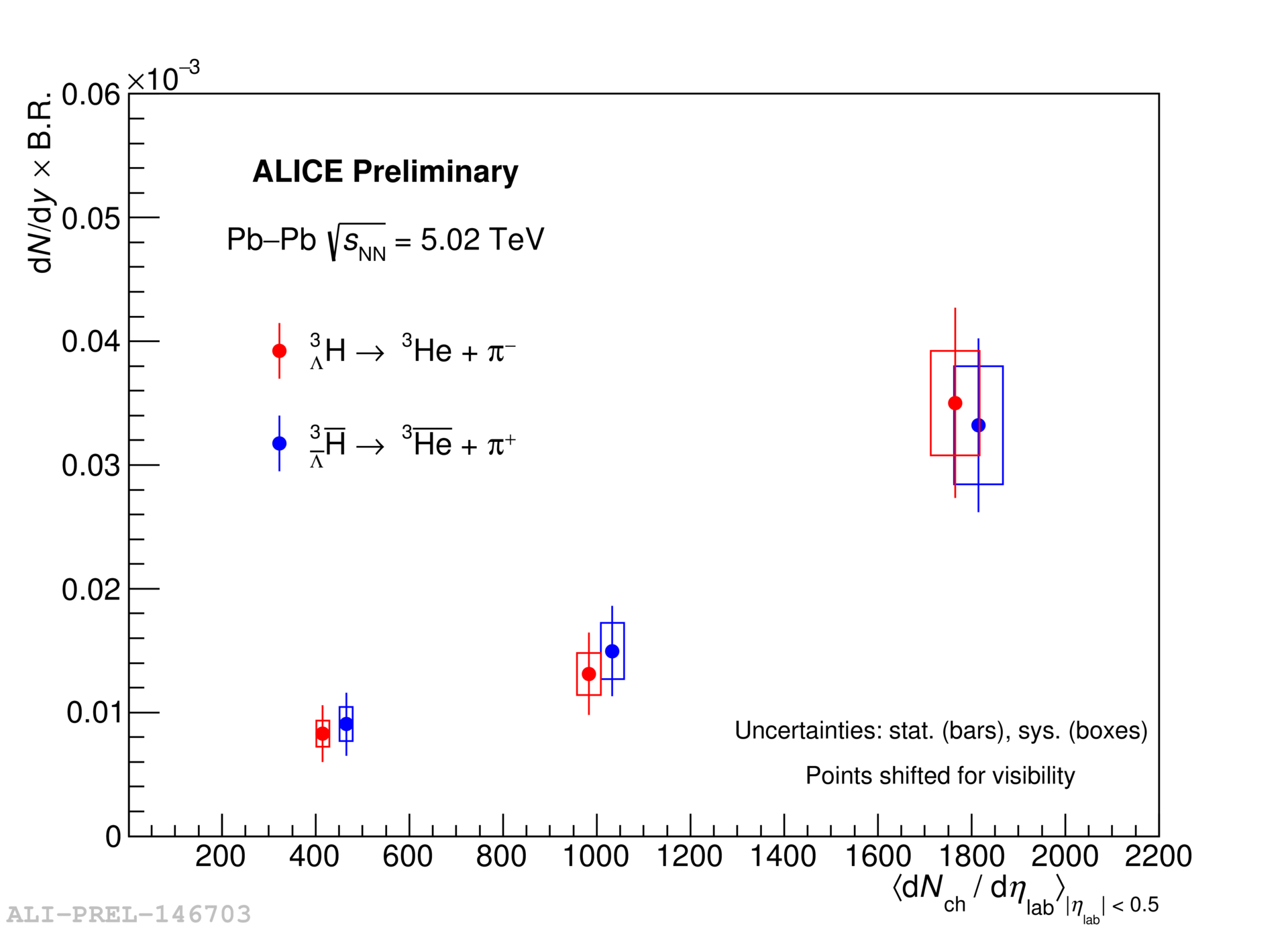}
\end{subfigure}%
\begin{subfigure}[h]{.5\textwidth}
  \centering
  \includegraphics[scale=0.095]{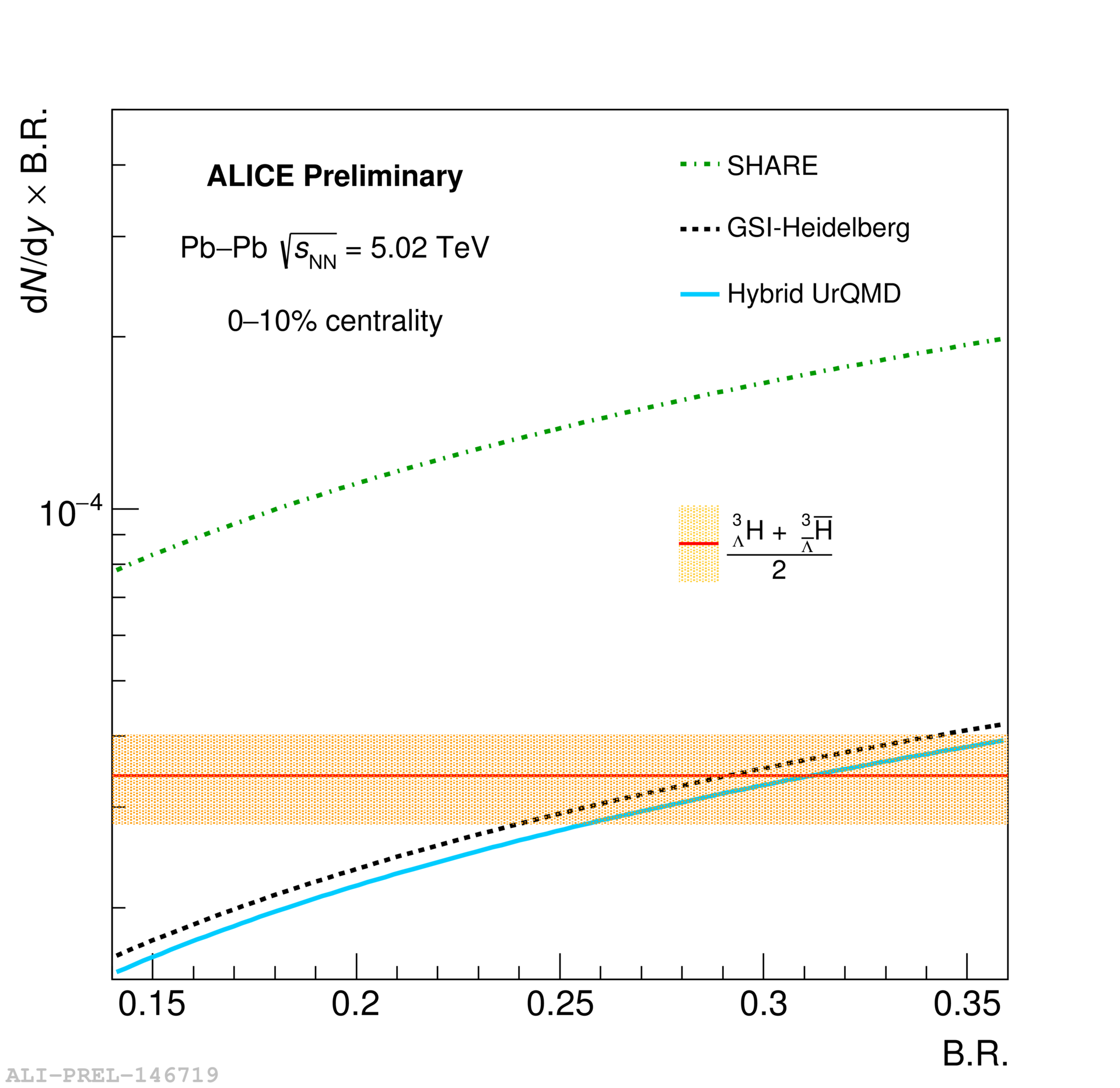}
\end{subfigure}
\caption{(\textit{left}) d$N$/d$y$ of $^3_{\Lambda}$H (red) and $^{3}_{\overline{\Lambda}}\mathrm{\overline{H}}$ (blue) as a function of the charged particle multiplicity d$N$/d$\eta$. Points shifted for visibility; (\textit{right}) averaged d$N$/d$y$ measured in 0-10$\%$ centrality class compared with thermal model predictions \cite{PLB697:2011,Coal_2,share2013} as a function of the decay branching ratio (B.R.) }
\label{dndy}
\end{figure}

The production yields d$N$/d$y$ have been measured for three different centrality classes (0-10$\%$, 10-30$\%$ and 30-50$\%$) 
Figure \ref{dndy} (\textit{left}) shows the corrected yields, d$N$/d$y$, evolution with the charged particle multiplicity d$N$/d$\eta$ for $^3_{\Lambda}$H (red) and $^{3}_{\overline{\Lambda}}\mathrm{\overline{H}}$ (blue). The increasing trend can be interpreted, according to thermal models, as related to the increasing size of the volume of the created medium.  
Since the decay branching ratio (B.R.) is barely known and constrained by the ratio between all charged channels containing a pion (R = $\Gamma^{^3\rm{He}}$/($\Gamma^{^3\rm{He}}$ + $\Gamma^{\rm{p+d}}$ + $\Gamma^{\rm{p+p+n}}$)) \cite{Kamada:1997rv}, the averaged d$N$/d$y$ measured in 0-10$\%$ is compared with thermal model predictions as a function of the B.R. as shown in Fig. \ref{dndy} (\textit{right}). The equilibrium thermal models, GSI-Heidelberg \cite{PLB697:2011} and Hybrid UrQMD \cite{Coal_2}, are in agreement with the measured yield in the B.R. range between 24$\%$ and 35$\%$, while the non-equilibrium model SHARE \cite{share2013} overestimates the yield by a factor 4 for the expected B.R. of 25$\%$ \cite{Kamada:1997rv}.

\begin{figure}[!h]
\centering
\begin{subfigure}[h]{.5\textwidth}
  \centering
  \includegraphics[scale=0.1]{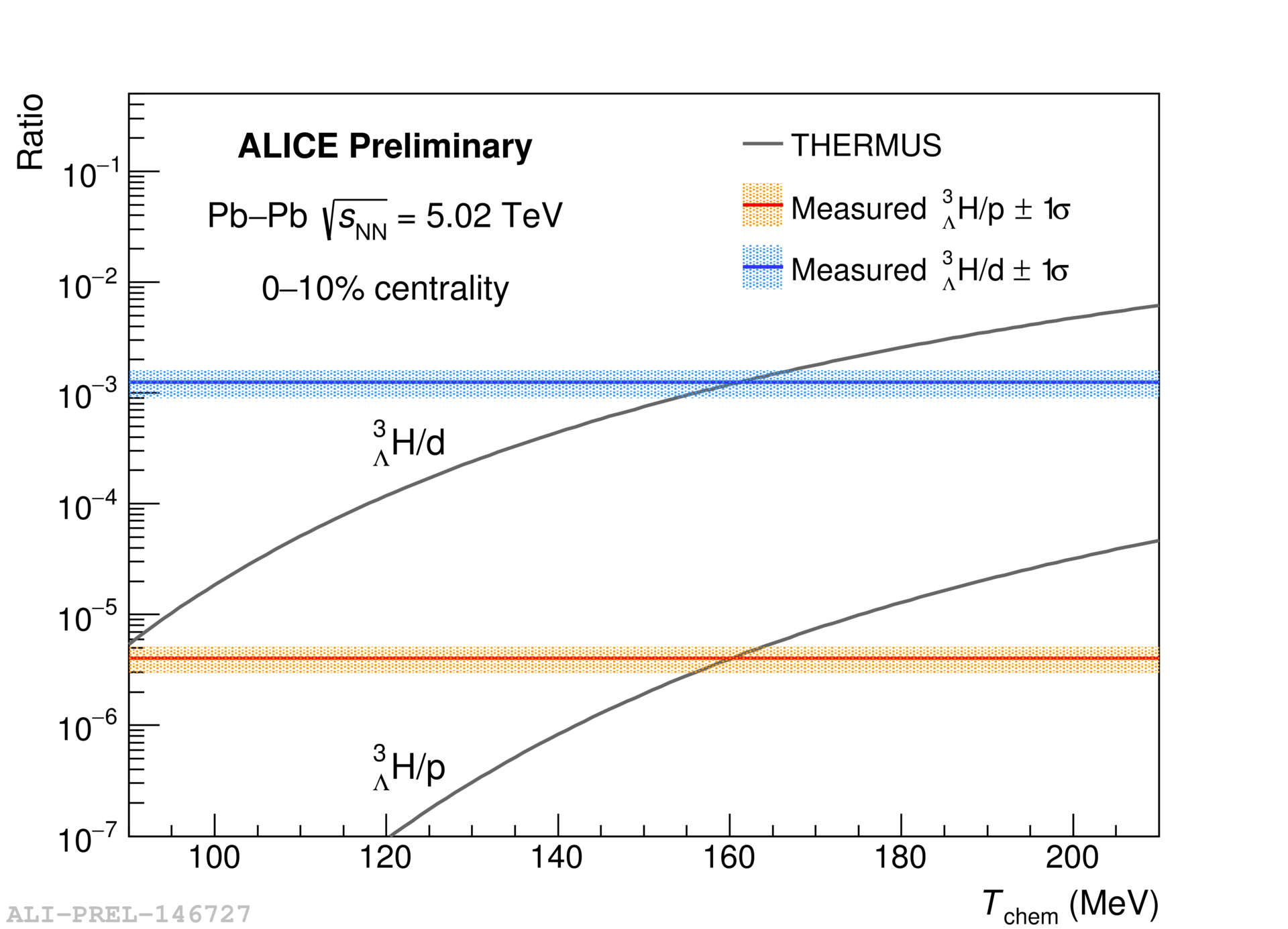}
\end{subfigure}%
\begin{subfigure}[h]{.5\textwidth}
  \centering
  \includegraphics[scale=0.095]{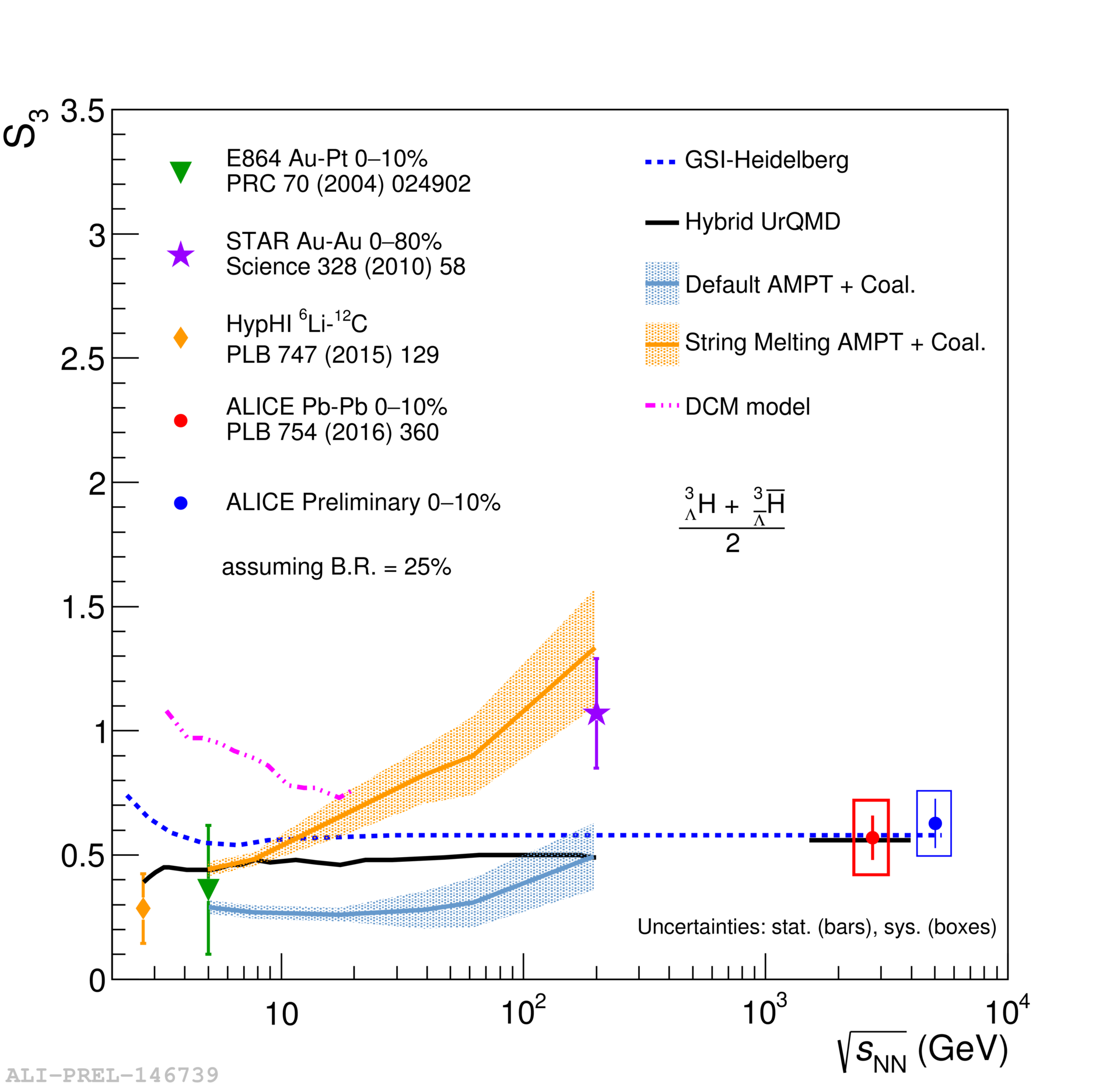}
\end{subfigure}
\caption{(\textit{left}) $^3_{\Lambda}$H/p (red) and $^3_{\Lambda}$H/d (blue) ratios compared with the THERMUS \cite{thermus2004} predictions as a function of T$_{chem}$ ; (\textit{right}) $S_3$ measured at different nucleon-nucleon centre-of-mass energies, with the ALICE preliminary result (blue circle), compared to thermal \cite{PLB697:2011, Coal_2} and coalescence \cite{Zhang:2009ba, ampt:2005ba} model predictions.}
\label{ratios}
\end{figure}

The ratio between $^3_{\Lambda}$H and proton or deuteron yields has been calculated since ratios to light hadrons yields are more sensitive to the chemical freeze-out temperature. Figure \ref{ratios} (\textit{left}) shows the ratios of $^3_{\Lambda}$H to proton (red) and to deuteron (blue) yields in the 0-10$\%$ centrality class. The coloured bands are the square roots of the sum in quadrature of the statistical and systematic uncertainties. These ratios are compared with the predictions of THERMUS \cite{thermus2004} as a function of T$_{chem}$ and they turned out to be in agreement with the model in the temperature range 153 and 165 MeV. 
 
The strangeness population factor $S_3$, defined as $S_3$ = $\frac{^3_{\Lambda}\rm H}{^3\mathrm{He}} \times \frac{p}{\Lambda}$ \cite{Zhang:2009ba}, is a double ratio independent of the chemical potential of the particles and any additional canonical correction factor for strangeness is cancelled. Figure \ref{ratios} (\textit{right}) shows $S_3$ measured by different experiments at various nucleon-nucleon centre-of-mass energies. 
The ALICE preliminary result (blue circle), obtained in Pb--Pb collisions at $\sqrt{s_{\mathrm{NN}}}$ = 5.02 TeV, is the average between $^3_{\Lambda}$H and $^{3}_{\overline{\Lambda}}\mathrm{\overline{H}}$ values and the statistical and systematic uncertainties are represented as a line and a box respectively. This result is in agreement with that obtained by analyzing the Pb--Pb collisions at $\sqrt{s_{\mathrm{NN}}}$ = 2.76 TeV \cite{Adam:2015yta}. 
The $S_3$ predictions from GSI-Heidelberg and Hybrid UrQMD at the LHC energies are in agreement with the measured value, while the predictions from coalescence models are available only up to top RHIC energy.

\begin{figure}[!h]
\centering
\includegraphics[scale=0.135]{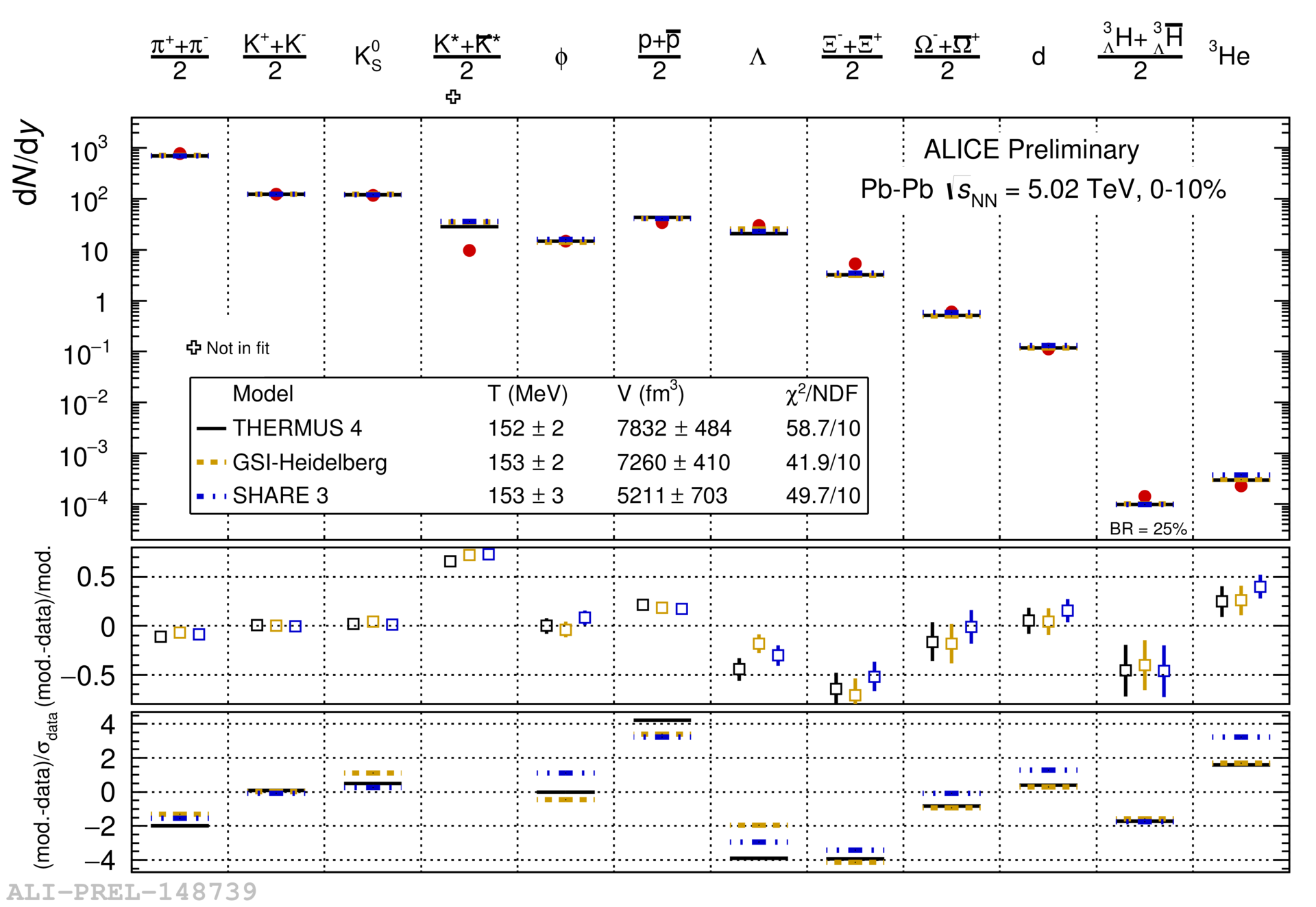}
\caption{Thermal model fits with three different implementations to the light flavour hadron yields in central (0-10$\%$) Pb-Pb collisions at $\sqrt{s_{\mathrm{NN}}}$ = 5.02 TeV.}
\label{th_fit}
\end{figure}

Finally, the $^3_{\Lambda}$H yield is included in the thermal fit performed to all particle production yields measured by the ALICE Collaboration in Pb--Pb collisions at $\sqrt{s_{\mathrm{NN}}}$ = 5.02 TeV in 0-10$\%$ centrality class. The fit was performed with three model implementations \cite{PLB697:2011, thermus2004, share2013}. The d$N$/d$y$ are qualitatively well described by these models with T$_{chem}$ = 153 $\pm$ 2 MeV assuming statistical-thermal equilibrium as shown in Fig. \ref{th_fit}. 

\section{Lifetime}
\label{life}
The $^3_{\Lambda}$H is a loosely bound object, with $\Lambda$ separation energy $B_{\Lambda}$ = 0.13 $\pm$ 0.05 MeV \cite{Davis}, and the theory predicts a value of the lifetime compatible with the free $\Lambda$ lifetime \cite{Kamada:1997rv}. However recent results from heavy-ion experiments are below the expected free $\Lambda$ lifetime.
ALICE measured a lifetime value, analyzing the Pb--Pb data sample at $\sqrt{s_{\mathrm{NN}}}$ = 2.76 TeV \cite{Adam:2015yta}, which is in agreement with this trend and the world average lifetime $\tau$ = 215$^{+18}_{-16}$ ps \cite{Rappold:2014} as shown in Fig. \ref{life} (\textit{left}).
\begin{figure}[!h]
\centering
\begin{subfigure}[h]{.5\textwidth}
  \centering
  \includegraphics[scale=0.42]{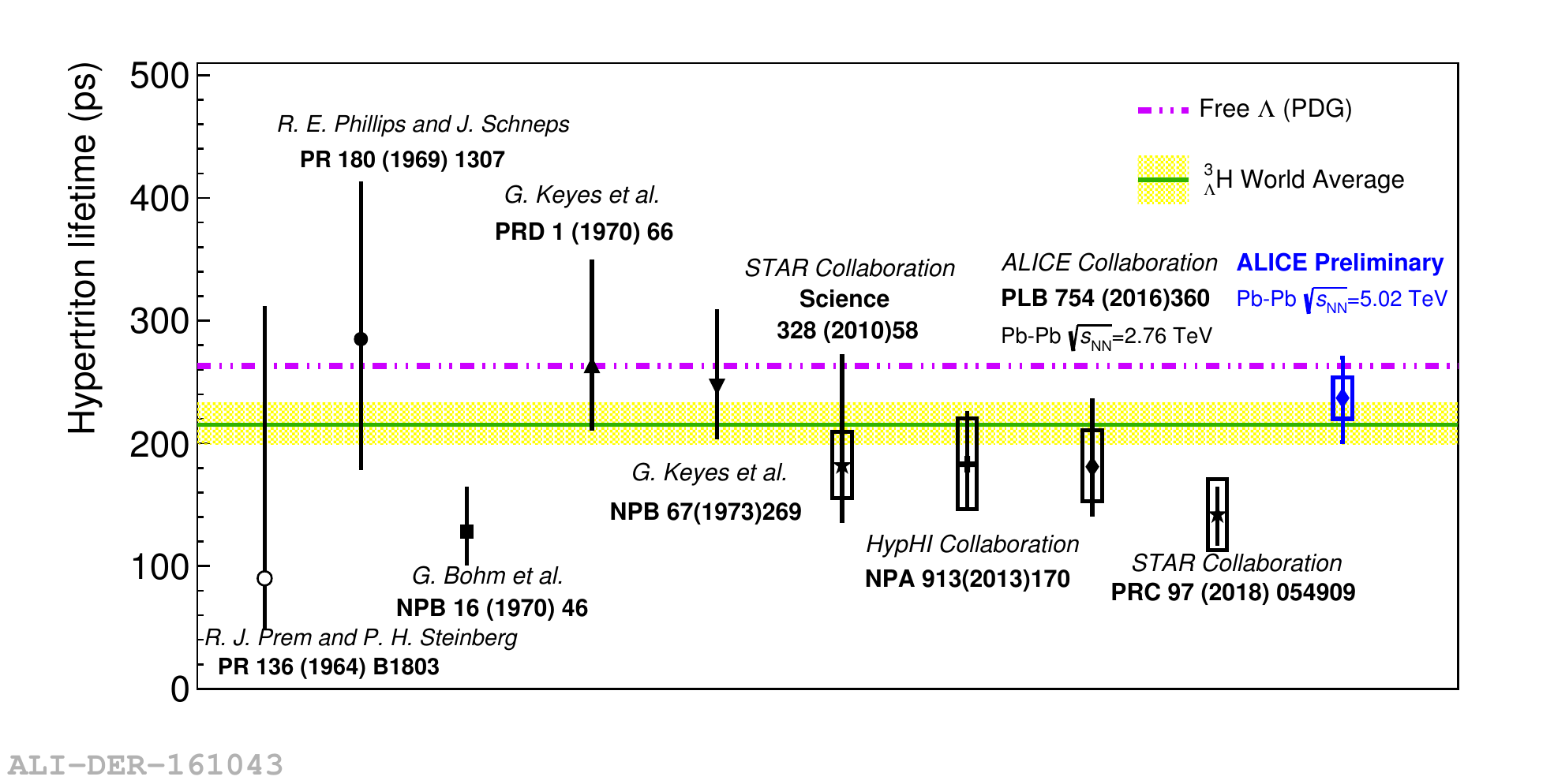}
\end{subfigure}%
\begin{subfigure}[h]{.5\textwidth}
  \centering
  \includegraphics[scale=0.26]{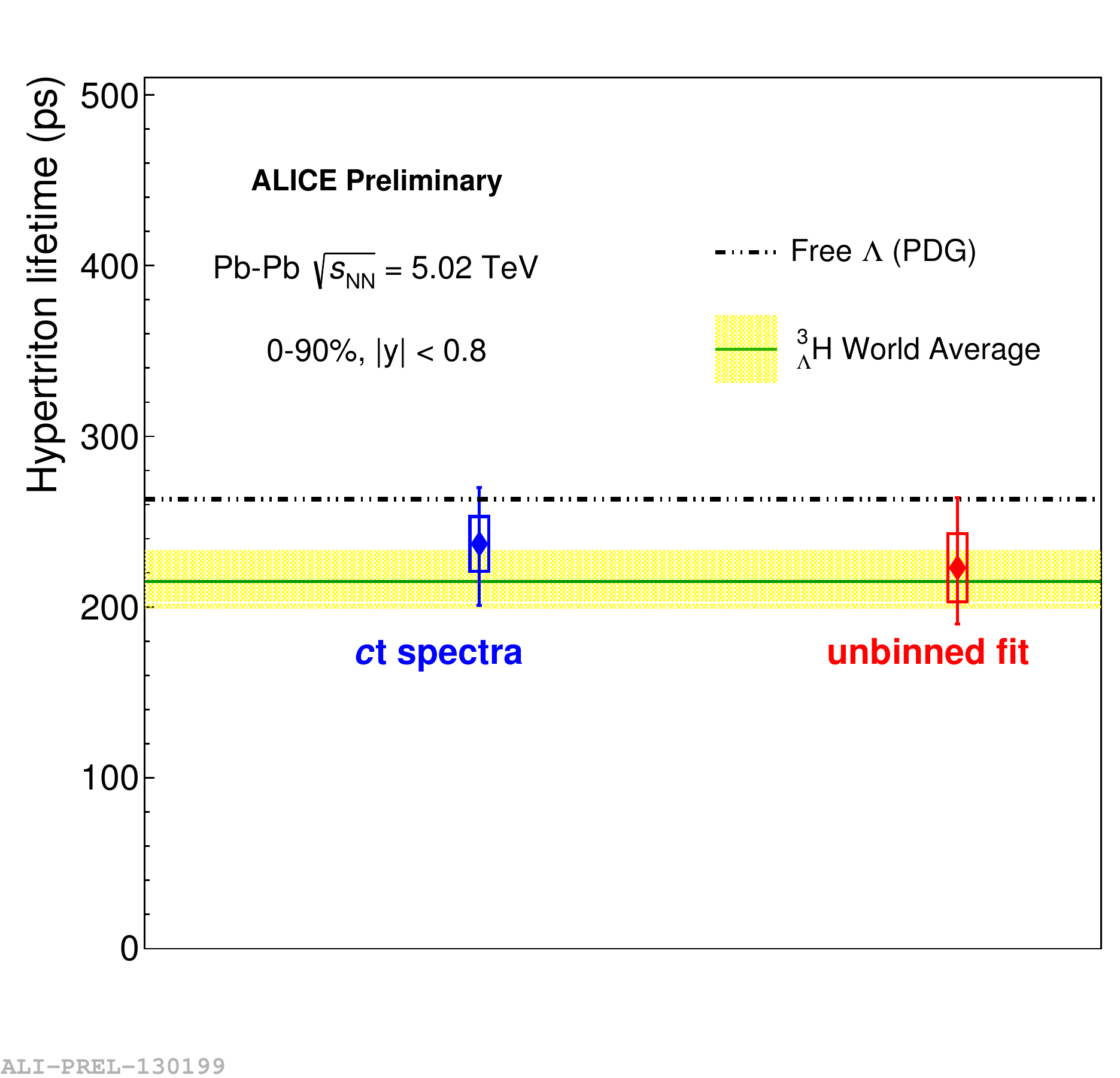}
\end{subfigure}
\caption{(\textit{left}) Lifetime values available in literature with the new ALICE result (blue marker) obtained analysing the data sample of Pb--Pb collisions at $\sqrt{s_{\mathrm{NN}}}$ = 5.02 TeV; (\textit{right}) $^3_{\Lambda}$H lifetime values obtained with the two methods.}
\label{life}
\end{figure}
The $^3_{\Lambda}$H lifetime has been measured with higher precision analysing the data sample of Pb--Pb collisions at $\sqrt{s_{\mathrm{NN}}}$=5.02 TeV. The decay products and the secondary vertices have been reconstructed following the same strategy adopted for the production yields.

Two techniques have been used for the lifetime determination. The first one consists of the exponential fit to the background subtracted $ct$-differential spectrum, leading to a lifetime value $\tau$ = 237$^{+33}_{-36}$(stat.) $\pm$ 17(syst.) ps, shown as blue marker in Fig. \ref{life} (\textit{left}). This figure shows a collection of the $\tau$ values, including the latest STAR result \cite{PhysRevC.97.054909}. 
The second approach is an unbinned fit to the 2-dimensional distribution $ct$ vs invariant mass. The analysis is performed with a first fit to the invariant mass distribution with the same function used in the signal extraction for the production analysis. The signal range ($\mu$ $\pm$ 3$\sigma$) and the sidebands ($\mu$ $\pm$ 3$\sigma$, $\mu$ $\pm$ 9$\sigma$) are defined using $\mu$ and $\sigma$ from the invariant mass fit.
The lifetime is determined with a fit to the $ct$ distribution in the signal range with an exponential function for the signal and a double exponential for the background, which has been tuned in the sidebands. The lifetime value obtained with this second method is in agreement within 1$\sigma$ with the one from the first method, as shown in Fig. \ref{life} (\textit{right}) .   

\section{Conclusion}
\label{conclusion}
The data sample of Pb--Pb collisions at $\sqrt{s_{\mathrm{NN}}}$ = 5.02 TeV, collected by the ALICE experiment, allowed to measure the production and to determine the lifetime of $^3_{\Lambda}$H and $^{3}_{\overline{\Lambda}}\mathrm{\overline{H}}$.
The preliminary results on production yields and spectra are well described by statistical-thermal models, assuming statistical equilibrium and a chemical freeze-out temperature around 153 MeV. 
On the other hand, the predictions from coalescence model are available only up to RHIC energies at the moment and need to be expanded to the LHC energies for a quantitative comparison.
The lifetime of the $^3_{\Lambda}$H has been measured with more data and better precision and the preliminary result is closer to the expected free $\Lambda$ lifetime. Further improvement on the lifetime is expected by exploiting the $^3_{\Lambda}$H 3-body decay channel. Future data taking in LHC Runs 2, 3 and 4 will allow a reduction of the statistical uncertainties on the lifetime determination. 




\bibliographystyle{elsarticle-num}
\bibliography{trogolo_bib}

\begin{thebibliography}{10}
\expandafter\ifx\csname url\endcsname\relax
  \def\url#1{\texttt{#1}}\fi
\expandafter\ifx\csname urlprefix\endcsname\relax\def\urlprefix{URL }\fi
\expandafter\ifx\csname href\endcsname\relax
  \def\href#1#2{#2} \def\path#1{#1}\fi

\bibitem{Davis}
D.~Davis, Nucl. Phys. A754 (2005) 3.

\bibitem{PLB697:2011}
A.~Andronic, P.~Braun-Munzinger, J.~Stachel, H.~Stocker, Phys. Lett. B697
  (2011) 203.

\bibitem{Coal_1}
J.~Kapusta, Phys. Rev. C21 (1980) 1301.

\bibitem{Coal_2}
J.~Steinheimer, K.~Gudima, A.~Botvina, I.~Mishustin, M.~Bleicher, H.~Stocker,
  Phys. Lett. B714 (2012) 85--91.

\bibitem{Congletone:1992jp}
J.~Congleton, J. Phys. G Nucl. Part. Phys. 18 (1992) 339.

\bibitem{Kamada:1997rv}
H.~Kamada, J.~Golak, K.~Miyagawa, H.~Witala, W.~Gloeckle, Phys. Rev. C57 (1998)
  1595.

\bibitem{share2013}
M.~Petr\'{a}\v{n}, J.~Letessier, V.~Petr\'{a}\v{c}ek, J.~Rafelski, Phys. Rev.
  C88~(3) (2013) 034907.

\bibitem{thermus2004}
S.~Wheaton, J.~Cleymans, M.~Hauer, Comput. Phys. Commun. 180 (2009) 84.

\bibitem{Zhang:2009ba}
S.~Zhang, J.~H. Chen, H.~Crawford, D.~Keane, Y.~G. Ma, Z.~B. Xu, Phys. Lett.
  B684 (2010) 224.

\bibitem{ampt:2005ba}
Z.-W. Lin, C.~Ko, B.-A. Li, B.~Zhang, S.~Pal, Phys. Rev. C72 (2005) 064901.

\bibitem{Adam:2015yta}
J.~Adam, et~al., Phys. Lett. B754 (2016) 360.

\bibitem{Rappold:2014}
C.~Rappold, et~al., Phys. Lett. B728 (2014) 543.

\bibitem{PhysRevC.97.054909}
L.~Adamczyk, et~al., Phys. Rev. C97 (2018) 054909.

\end{thebibliography}







\end{document}